\documentclass[12 pt]{article}

\usepackage{amssymb,amsfonts,latexsym}

\begin{document}

\newtheorem{dfn}{Definition}[section]
\newtheorem{theorem}{Theorem}[section]
\newtheorem{axiom2}{Example}[section]
\newtheorem{axiom3}{Lemma}[section]
\newtheorem{lem}{Lemma}[section]
\newtheorem{prop}{Proposition}[section]
\newtheorem{cor}{Corollary}[section]
\newcommand{\be}{\begin{equation}}
\newcommand{\ee}{\end{equation}}
\newcommand{\lmat}{\left(\begin{array}{cccccc}}
\newcommand{\rmat}{\end{array}\right)}
\newcommand{\lm}{\lambda}
\newcommand{\al}{\alpha}
\newcommand{\ID}{{\mathbb{D}}}
\newcommand{\IG}{{\mathbb{G}}}
\newcommand{\X}{{\mathbb{X}}}
\newcommand{\Y}{{\mathbb{Y}}}
\newcommand{\p}{\partial}
\title{Exploring Branched Hamiltonians For A Class Of Nonlinear Systems\\}

\author {Bijan Bagchi\footnote{E-mail: {\tt bbagchi123@gmail.com}}\\
Department of Applied Mathematics, University of Calcutta\\
 92 Acharya Prafulla Chandra Road \\ Kolkata - 700009,  India \\
\and
Subhrajit Modak and
Prasanta K. Panigrahi
 \footnote{E-mail: {\tt modoksuvrojit@gmail.com, panigrahi.iiser@gmail.com}}\\
Indian Institute of Science,\\
 Education and Research (Kolkata), \\ Mohanpur, West Bengal 741 246, India.\\
\and
Franti\v{s}ek Ruzicka and Miloslav Znojil
\footnote{E-mail: {\tt fruzicka@gmail.com,
znojil@ujf.cas.cz}}\\
Nuclear Physics Institute ASCR, 250 68 $\check{R}$e$\check{z}$
\\ Czech Republic \\}

\date{}

\maketitle

\smallskip

\smallskip

\begin{abstract}

One of the less well understood ambiguities of quantization is
emphasized to result from the presence of higher-order time
derivatives in the Lagrangians resulting in multiple-valued
Hamiltonians. We explore certain classes of branched Hamiltonians in
the context of nonlinear autonomous differential equation of
Li\'{e}nard type. Two eligible elementary nonlinear models that
emerge are shown to admit a feasible quantization along these lines.

 \end{abstract}

\paragraph{Mathematical Classification} 34C14, 70H33.

\smallskip


\section{Introduction}

The Bohr's formulation of his model of atom \cite{Bohr} was one of
the most important steps towards the subsequent birth of
full-fledged quantum theory from classical mechanics. His choice of
benchmark systems was insightful (quantum hydrogen with Coulomb
potential $V(\vec{x})\sim 1/|\vec{x}|$ in non-relativistic
Schr\"{o}dinger equation is one of not too may exactly solvable
real-world systems) and lucky (indeed, the subsequent inclusion of
subtleties led just to some not too essential corrections and
modifications of the picture).

Today, more than 100 years later, the Bohr's emphasis on
connectedness between classical and quantum worlds is still alive
and inspiring. Although we are currently paying attention to
multiple much more sophisticated quantum systems, the principle of
correspondence and the traditional description of dynamics via
certain real and local potentials $V(\vec{x})$ still guides the
majority of phenomenological studies in atomic, molecular and
nuclear physics.

In a broader methodical framework, say, of condensed matter physics,
the traditional approach does not work that well. Incessantly, its
applicability is being challenged by theoretical innovations. Just
to name a few, let us mention the recent growth of popularity of the
detailed study of certain non-real, complex potentials with real
spectra \cite{book} or, most recently, of their friendly and
tractable non-local generalizations \cite{zno}. An
emerging interest also involves various non-linear forms of the
interactions which may be mediated, say, by an energy-dependence of
the dynamical parameters \cite{gezian}. Last but not least, let us
mention the fruitful transfer of the dynamics-determining role from
elementary potentials $V(x)$ to the other factors like, e.g.,
coordinate-dependent masses $m=m(x)$ \cite{qus, bij1, bij2} etc.

In this context a new class of innovations of the description and
simulation of quantum dynamics recently emerged in connection with
the possible specific role of the models of physical systems in
which the Lagrangians possess the time derivatives in excess of
quadratic powers \cite{asha,ashap,tlcu,tlcur,tlcurt,ros}. The use of
these models leads, on both the classical and quantum level, to the
necessity of a re-evaluation of the dynamical interpretation of the
momentum $p$ which, in principle, becomes a multiple function of
velocity $v$. In what follows we intend to present a few results in
this direction.

\section{Branched Hamiltonians}

A typical classical model of the above-mentioned non-quadratic type
may be sampled by Lagrangian
\begin{equation}\label{one}
L = (v - 1)^{\frac{2k-1}{2k+1}}- V(x)\,.
\end{equation}
Function $V(x)$ represents a convenient local interaction potential
while the traditional kinetic-energy term is tentatively replaced by
a fairly unusual function of ``velocity'' $v$.

This model was recently analyzed in Ref.~\cite{tlcurtr} where the
definition of the fractional powers of difference $v-1$ was adapted
to the needs of possible phenomenology. In detail, the $(2k+1)-$st
root was required real and positive or negative for $v>1$ or $v<1$,
respectively. Correspondingly, the quantity $v$ turned out to be a
double-valued function of $p$.

\subsection{The problem of quantization}

In a search for a quantum version or analogue of model (\ref{one}),
both of its latter features appear truly challenging. Firstly, the
definition using multivalued functions (referring to the literature
let us speak about ``branched Hamiltonians'') seems to require a
deeper insight in its physical nature and connections.

In this note we decided to consider the subject of the branched
Hamiltonians in the context of a class of nonlinear autonomous
differential equations of Li\'{e}nard type \cite{Lin1,Lin2} since
these systems are of potential importance not only in optics
\cite{DW} but also in the shallow-water-wave studies \cite{ZSL} and
in non-Hermitian quantum mechanics \cite{bag}. Due to nonlinearity
one of the main obstacles lies in the absence of superposition
principle. The recent progress in the direction of seeking
analytical and numerical solutions as well as of tracking down new
mathematical properties has been dramatic, nevertheless. In
particular, an inspiring feature of non-linear oscillations was
found in the dependence of amplitudes on the frequency, etc. A
useful role in this analysis was played, e.g., by the study of the
class of Li\'{e}nard equations
 \begin{equation}\label{two}
\ddot{x} + r(x)\dot{x} + q(x) = 0
\end{equation}
where overdot indicates a derivative with respect to the time
variable and where $r(x)$ and $q(x)$ are two continuously
differentiable functions of the spatial coordinate $x$.

After a restriction of the model to the specific form of $r(x)=kx$
and $q(x)=\lambda x + \frac{k^{2}}{9}x^{3}$ (which are odd functions
of $x$) one obtains

 \begin{equation}\label{three}
 \ddot{x} + kx\dot{x} + \frac{k^{2}}{9}x^{3}
 + \lambda x = 0, \quad \lambda > 0\,.
 \end{equation}
This equation represents a cubic oscillator subject to a damped
nonlinear force as indicated by the product $kx\dot{x}$. Naturally,
the presence of the damping is a challenge whenever one tries to
contemplate a quantization of the model.

\subsection{Hamiltonians}

The classically damped systems leading to non-Hermitian quantum
Hamiltonians need not necessarily imply a quantum damping
\cite{Gong}. Moreover, a backward classical-optics reinterpretation
of quantum Hamiltonians of this type may exist in photonics playing
the role of so called lossy amplifiers \cite{Ge}. Last but not
least, the similar classical-physics contexts (related, e.g., to the
study of Bose-Einstein condensates \cite{Dast, Dohnal}) re-open the
phenomenological appeal of classical nonlinearities as sampled by
Eq.~(\ref{one}). For all of these reasons it is worth checking the
conversion of the Lagrangians of  Eq.~(\ref{three}) into
Hamiltonians.

A brief account of some background and references is in order. First
of all, let us recollect \cite{vchi} that the toy-model Lagrangians
of the form

\begin{equation}\label{four}
L = \frac{27\lambda^{3}}{2k^{2}}(k\dot{x}+\frac{k^{2}x^{2}}{3}
+3\lambda)^{-1}+\frac{3\lambda\dot{x}}{2k} - \frac{9\lambda^{2}}{2k^{2}}
\end{equation}
admit the evaluation of the canonically conjugate momentum,

 \begin{equation}\label{five}
 p = -\frac{27\lambda^{3}}{2k}(k\dot{x}+\frac{k^{2}x^{2}}{3}
 +3\lambda)^{2}+\frac{3\lambda}{2k}\,.
 \end{equation}
Thus, the Hamiltonian can be cast in compact form

 \begin{equation}\label{6}
 H_{(x,p)}=\frac{9\lambda^{2}}{2k^{2}}
 \left[2-2\left(1-\frac{2kp}{3\lambda}
 \right)^{\frac{1}{2}}
 +\frac{k^{2}x^{2}}{9\lambda}
 -\frac{2kp}{3\lambda}-\frac{2k^{3}x^{2}p}{27\lambda^{2}}\right]\,.
\end{equation}
Such a Hamiltonian is of non-standard type. The co-ordinates and
potentials are mixed so that the expression cannot be split into
individual kinetic and potential energy terms. However, we can write

 \begin{equation}\label{7}
 H_{(x,p)}= \frac{1}{2}f(p)x^{2} + U(p)
 \end{equation}
where

 \begin{equation}\label{8}
 f(p) = \omega^{2} (1-\frac{2kp}{3\omega^{2}}), \quad  U(p)=
 \frac{9\omega^{4}}{2k^{2}}(\sqrt{1-\frac{2kp}{3\omega^{2}}}-1)^{2}
 \end{equation}
with $\omega = \sqrt{\lambda}$. The roles of coordinate and momentum
have been transposed implying a momentum-dependent system at play.
The first term in (\ref{8}) represents a mixed function of both
position and momentum variables while the second term is a function
of momentum alone. The classical Hamiltonian $H_{(x,p)}$ is
invariant under a joint action of coordinate reflection and time
reversal transformation.

Exact quantization of the Hamiltonian can be carried out by going
over to the momentum space with
$\hat{x}=i\hbar\frac{\partial}{\partial p}$. The Hamiltonian turns
out to be of momentum-dependent mass type. Adopting a von Roos
strategy of quantizing the problem by considering a general
symmetric ordering \cite{ovon}, the underlying Schrodinger equation
can be formulated and solved, in principle at least.

\section{Two illustrative models of dynamics}

In practice, the price to pay for having the model quantized lies in
the clarification of the role of singularities in the eigenfunction.
In this sense the road to quantization is, expectedly, not free of
difficulties. One can, however, feel encouraged by the work by
Chithiika Ruby et al \cite{vchi} who showed that the energy spectrum
and normalized solutions could still be obtained for a class of
Hamiltonians that are nonsymmetric and non-Hermitian.

Needless to add that the non-Hermitian nature of quantum Hamiltonian
may bring about a number of unpleasant consequences as, for example,
the emergence of the exceptional points \cite{Baga} or the breakdown
of the adiabatic theorem \cite{Hash, Rotter}. At the same time, the
acceptance of the anomaly may prove innovative, e.g., by giving a
new physical interpretation to wave packets \cite{Graefe} or to
pseudospectra \cite{Novak, Ruz}.

\subsection{An amendment of the Lagrangian}

In a way inspired by Eq.~(\ref{one}) let us consider a higher-power
Lagrangian
\begin{equation}\label{9}
L = C(v + f(x))^{\frac{2m+1}{2m-1}}-\delta,\quad C=
 \left(\frac{1-2m}{1+2m}\right)(\delta)^{\frac{2}{1-2m}},
 \quad \delta> 0.
\end{equation}
The main difference from (\ref{one}) lies in our choice of a general
function $f(x)$ in place of $f(x)=-1$ in our initial $L$. The other point is that the inverse exponent with respect to the model of Curtright and Zachos \cite{tlcurtr} is done for convenience of calculus. Further,
we have omitted an explicit presence of the potential function
assuming that the interaction re-appears in a more natural manner
via a suitable choice of an auxiliary free parameter $\delta$ and of a nontrivial function $f(x)$.  As long as our Lagrangian $L$ is of
a nonstandard type, we will not feel disturbed by the absence of the
explicit potential  $V(x)$.

Parameter $C$ is non-negative for $0 \leq m < \frac{1}{2}$ and the
canonical momentum is given by formula
 \begin{equation}\label{10}
 p=\frac{\partial L}{\partial v}
 = -(\delta)^{\frac{2}{1-2m}}(v+f(x))^{\frac{2}{2m-1}}
 \end{equation}
which can easily be inverted to yield
 \begin{equation}\label{11}
 v=-f(x)+\delta(\pm\sqrt{-p)}^{2m-1}\,.
 \end{equation}
The Hamiltonian has the structure
 \begin{equation}\label{12}
 H_{\pm}(x,p)=(-p) f(x)-
\frac{2\delta}{{2m+1}}(\pm\sqrt{-p})^{2m+1}  \quad + \delta\,.
 \end{equation}
For the present purposes it is worthwhile to inquire into the
specific case with $m=0$.

\subsection{Type I Hamiltonian at $m=0$}

At $m=0$ we easily derive the double-valued
\begin{equation}\label{13}
v=v_{\pm}= -f(x) \pm \frac{\delta}{\sqrt{-p}}\,.
\end{equation}
The  Hamiltonian branches out to the components
\begin{equation}\label{14}
H_{\pm}(x,p)= (-p)f(x)\mp{2\delta}\sqrt{-p}+\delta\,.
\end{equation}
It is readily noticed that the real or the complex character of
$H_{\pm}$ depends on the sign of the momentum $p$. Once we specify
 \begin{equation}\label{15}
 f(x)=\frac{\lambda}{2}x^{2}
 +\frac{9\lambda^{2}}{2k^{2}},
 \quad \delta = \frac{9\lambda^{2}}{2k^{2}}, \quad \lambda > 0
 \end{equation}
then under a shift $p\rightarrow \frac{2k}{3\lambda}p- 1$, $H_{\pm}$
move over to the corresponding forms
 \begin{equation}\label{16}
 H_{\pm}(x,p)=\frac{9\lambda^{2}}{2k^{2}}
 \left[2  \mp 2\left(1-\frac{2kp}{3\lambda}
 \right)^{\frac{1}{2}}
 +\frac{k^{2}x^{2}}{9\lambda}
 -\frac{2kp}{3\lambda}-\frac{2k^{3}x^{2}p}{27\lambda^{2}}\right]\,.
 \end{equation}
These represent a set of plausible Hamiltonians for the nonlinear
system (\ref{three}) possessing coordinate-momentum coupling terms.
While $H_{+}$ coincides with the form of $H_{(x,p)}$ as furnished in
(\ref{6}), in both $H_{\pm}$ the presence of a linear harmonic
oscillator potential is revealed in the limit $k\rightarrow 0$.
The specific choice of the parameter $\delta$ as given above allows equation (\ref{three}) to be converted to a harmonic oscillator form under the nonlocal
transformation $U=xe^{\frac{k}{3}\int x(\tau)d\tau}$ \cite{vkch}.

In the classical scenario one needs to restrict $p$ to
$-\infty<p\leq\frac{3\omega^{2}}{2k}$  in order to address the
physical properties of the system in the real space.  However the
presence of a branch point singularity at $p=\frac{3\lambda}{2k}$
makes the study of $H_{\pm}$  quite complicated. At the boundary
$p=\frac{3\lambda}{2k}$, the two Hamiltonians $H_{\pm}$ coincide.

\subsection{Type II Hamiltonian}

We next focus on a class of models under the influence of the
Lagrangian
\begin{equation}\label{17}
L=\frac{1}{s}\left(\frac{1}{3}sx^{2}+\frac{3}{s}\lambda - v\right)^{-1}
\end{equation}
where $s$ is a real parameter. This implies, for the canonical
momentum, the result
 \begin{equation}\label{18}
 p=\frac{\partial L}{\partial v}=
 \frac{1}{s}\left(\frac{1}{3}sx^{2}+\frac{3}{s}\lambda -v\right)^{-2}\,.
 \end{equation}
Inversion of (18) leads to
 \begin{equation}\label{19}
 v= \frac{1}{3}sx^{2}+\frac{3}{s}\lambda \pm \frac{1}{\sqrt{sp}}\,.
 \end{equation}
Finally, employing the Legendre transform, the accompanying
Hamiltonian to the above Lagrangian can be projected, for the
two-$v$ branches, as
 \begin{equation}\label{20}
 H_{\pm}(x,p)=pv_{\pm}(p)-L=\frac{s}{3}x^{2}p
 +\frac{3}{s}\lambda p\mp2\sqrt{\frac{p}{s}}\,.
 \end{equation}
Eq.~(20) describes a set of alternative Hamiltonians for (3). The
component $H_{+}$ for the case $s<0$ has earlier been derived in
\cite{vkch} and found to possess an interesting $\lambda \to 0$
limit.

\section{Quantization}

\subsection{Schr\"{o}dinger equation}

Quantization of the branched system (\ref{16}) can be handled in a
typical way as detailed in \cite{vchi}, i.e., by adopting a suitable
ordering procedure and implementing appropriate boundary conditions.
In order to carry out quantization for the Type II Hamiltonian we
will consider here the case of $s>0$ and subject it to a
perturbative treatment. In this regard we will study the spectrum of
Hamiltonians entering Schr\"{o}dinger equation
 \begin{equation}\label{21}
 H_{\pm}(x,p)\,\psi^{(\pm)}(p)= \eta^{(\pm)}\,\psi^{(\pm)}(p)
 \end{equation}
where, in units $\hbar=1$,
 \begin{equation}\label{22}
 \frac{3}{s} H_{\pm}(x,p)/p=
 -\frac{d^2}{dp^2} \mp 6\sqrt{\frac{1}{s^3p}}+\frac{9}{s^2}\lambda\,.
 \end{equation}
After the change of variables $p=r^\varrho$ (with special
$\varrho=2$) and having set $p\psi^{(\pm)}(p)=r^\xi \chi(r)$ (with
special $\xi= \varrho+\frac{1}{2}$) one ends up with the following
transformed Schr\"{o}dinger equation
 \begin{equation}\label{23}
  -\chi''(r)+\frac{\ell(\ell+1)}{r^2}\,\chi(r) + \frac{36}{s^2}\,r^2\,\chi(r)
  \mp \frac{24}{s^{3/2}} \,r\,\chi(r)
  = \frac{12}{s}\,\eta^{(\pm)}\,\chi(r)
 \end{equation}
where the emergence of the centrifugal term is a byproduct of the change of variables \cite{class} and
where the quantity resembling the angular momentum is equal to $\ell=1/2$ for our choice of
exponent $\varrho = 2$.

\subsection{Perturbation expansions}

Eigenvalue problem (\ref{23}) is well suited for application of
Rayleigh-Schr\"{o}dinger perturbation theory \cite{BJ}. Towards its
implementation we abbreviate $\mp \frac{24}{s^{3/2}} \,r\,= g\,V$.
Assuming here that the value of $s$ is sufficiently large, the
auxiliary variable $g$ may be treated as a small parameter.
Hamiltonians $H=H^{(\pm)}$ as well as the related infinite sets of
wave-function kets $|\chi\rangle=|\chi^{(\pm)}_{(n)} \rangle$ and of
the bound-state energies $E=E^{(\pm)}_{(n)}$ may be then formally
Taylor-expanded,
 \begin{equation}\label{24}
 \label{ginpert}
 H=H^{(HO)} + g\,V\,,
 \ \ \ \
 |\chi \rangle = \sum_{m=0}^\infty\,g^m\,|\chi_m \rangle\,,
 \ \ \ \
 E = \sum_{m=0}^\infty\,g^m\,E_m\,.
 \end{equation}
Recalling also that the unperturbed Hamiltonian $H^{(HO)}$ stands
for the $\ell-$wave harmonic oscillator living on half-line of $r
\in (0,\infty)$ we arrive at the well known formulae for the
zero-order perturbation energies $E_0=\omega\,(4n+2\ell+3)$ where
$\omega=6s^{-1}$. The unperturbed eigenfunctions may also be
expressed in the compact form
 \begin{equation}\label{25}
 \langle r|\chi_0 \rangle={\cal N}\,
 r^{\ell+1}\exp (-\,\omega\,r^2/2)\,L_{n}^{(\alpha)}(\omega\,r)\,
 \end{equation}
containing a suitable normalization constant ${\cal N}$ and associated Laguerre
polynomials $L_{n}^{(\alpha)}(\omega\,r)\,$ with, in our present
special case, $\alpha=\ell+1/2=1\,$.

At this moment one would have to return to our hypothesis of
smallness of $g$ and to prove that the infinite-series ansatz of
eq.~(\ref{24}) are all convergent whenever $|g|\leq R$, i.e.,
at some non-vanishing radius of convergence $R>0$. Fortunately the proof is
an easy consequence of the fact that our operator of
perturbation $V$ is relatively bounded with respect to $H^{(HO)}$.

In the final step of the construction of bound states we only have
to insert formulae~(\ref{24}) in Schr\"{o}dinger equation, deduce
the recurrent Rayleigh-Schr\"{o}dinger perturbation recipe and write
down the corresponding general formulae that yield, typically,
 \begin{equation}\label{26}
 \eta^{(\pm)}= \eta^{(\pm)}_{(n)}= 2n+2 \mp \eta_{(n)}\,
 \ \ \ \ \  n=0,1,\ldots\,
 \end{equation}
where the quantities $\eta_{(n)}$ must be evaluated numerically.

\section{Summary}

Although virtually all of the available quantization recipes are
ambiguous, one can still restrict one's attention to certain
sufficiently elementary classical scenarios and succeed in the
construction of a mathematically consistent quantum model the
practical acceptability of which is to be verified via its
appropriate experimental realization.

In such a setting the key motivation of our present study emerged
after we imagined that the classical double-valued Hamiltonians
could offer a unique possibility of moving over to the quantum
sector thus opening up branched partners in such a setting too. We
decided to follow this idea and to perform the first steps in this
direction.

Our present results should be read as encouraging, confirming the
purely technical feasibility of the general strategy we described.
During its application we revealed that there exist several
nontrivial examples of the branched Hamiltonians for which the
quantization appeared not only feasible but also comparatively
friendly.

Beyond this framework, our present approach also seems to open an
entirely new direction of research in which one might be able to
combine the generic numerical techniques with the non-numerical
insight provided by the mathematics of multivalued functions in
combination with the analytic tools of perturbation theory.

In the future research we expect that our present approach could
also open a way towards the emergence of new qualitative features of
branched Hamiltonians. In particular, we expect that the new classes
of doublets of supersymmetric-partner Hamiltonians could emerge here
in some exotic manner and in a way which has not yet been followed
in the literature (cf., e.g., \cite{susy1, susy2, susy3, susy4}).

\section{Acknowledgements}

One of us (BB) thanks Anindya Ghose Choudhury and Partha Guha for discussions. We also thank the referee for constructive suggestions.


\end{document}